\documentclass[a4paper]{aa}  

\usepackage{graphicx}
\usepackage{subcaption}
\usepackage{txfonts}
\usepackage{lscape}
\usepackage{hyperref}
\usepackage{microtype}

\usepackage[dvipsnames]{xcolor}

\begin{document} 

\title{Variable, circularly polarized radio emission from the Young Stellar Object [BHB2007]-1: another ingredient of a unique system}
\authorrunning{S. Kaur et al.}
\titlerunning{[BHB2007]-1 circular polarized radio emission}
\author{Simranpreet Kaur \inst{1,2}, Josep M. Girart \inst{1,2},    Daniele Vigan\`o   \inst{1,2,3}, \'Alvaro S\'anchez Monge \inst{1,2}, L. Ilsedore Cleeves \inst{4}, Alice Zurlo\inst{5,6}, Fabio Del Sordo\inst{1,2,7}, \`Oscar Morata\inst{1}, Trisha Bhowmik\inst{5,6},
Felipe O. Alves\inst{8}}

\institute{Institut de Ci\`encies de I'Espai (ICE-CSIC), Campus UAB, Carrer de Can Magrans s/n, E-08193 Barcelona, Catalonia, Spain
\and
Institut d’Estudis Espacials de Catalunya (IEEC), E-08860 Barcelona, Catalonia, Spain
\and
Institute of Applied Computing \& Community Code (IAC3), Universitat de les Illes Balears, 
Palma, E-07122, Spain
\and
Department of Astronomy, University of Virginia, Charlottesville, VA 22904, USA
\and 
Instituto de Estudios Astrof\'isicos, Facultad de Ingenier\'ia y Ciencias, Universidad Diego Portales, Av. Ej\'ercito Libertador 441, Santiago, Chile
\and
Millennium Nucleus on Young Exoplanets and their Moons (YEMS) 
\and
INAF, Osservatorio Astrofisico di Catania, via Santa Sofia, 78 Catania, Italy
\and
House of Communication, Friedenstr. 24, Munich, Germany
\\
\email{simranpreet.k@csic.es}
}
  
\date{Received 02 May 2024 / Accepted 06 August 2024}

\abstract
{The young stellar object [BHB2007]-1 has been extensively studied in the past at radio, millimeter, and infrared wavelengths. It shows a gap in the disk and previous observations claimed the possible emission from a forming sub-stellar object, in correspondence to the disk gap. Here, we analyze a set of 8 \textit{Karl Jansky} Very Large Array (VLA) observations at 15 GHz and spread over a month. We infer a slowly variable emission from the star, with a $\sim 15-20\%$ circular polarization detected in two of the eight observations. The latter can be related to the magnetic fields in the system, while the unpolarized and moderately varying component can be indicative of free-free emission associated with jet induced shocks or interaction of the stellar wind with dense surrounding material. We discard any relevant short flaring activities when sampling the radio light curves down to 10 seconds and find no clear evidence of emission from the sub-stellar object inferred from past observations, although deeper observations could shed further light on this.}

\keywords{Protoplanetary disks; Stars: formation; Radio continuum: planetary systems; Polarization}

\maketitle
%

\section{Introduction}


Young stellar objects (YSOs), often found in dusty regions of the universe, are key to understanding star formation and the development of planetary systems. These objects, which emerge from clouds of gas and dust, are highly active and emit radiation across a broad electromagnetic spectrum, from X-rays to radio waves \citep{flares_2,flares,YSO_3,thermal_emission,YSO_5,YSO_6}. This broad emission spectrum makes them excellent subjects for studies using different wavelengths, especially radio and millimeter interferometry as such studies can provide important insights into the magnetic environments near these young stars \citep{YSO_3}, their flaring activities \citep{flares,flares_2}, and the characteristics of the material surrounding them. It also helps to identify early stages of star formation when the object is still deeply embedded in its original dust cloud \citep{thermal_emission}.

A significant contribution to the YSO emission is thermal in nature, at both radio and millimeter wavelengths. At millimeter wavelengths, this is mainly dominated by thermal dust emission 
\citep{thermal_emission,YSO_6},
while at centimeter wavelengths, the prime contributor is bremsstrahlung radiation in ionized gas \citep{thermal_emission,YSO_6}. In case of high-mass YSOs, this ionized gas is produced as a result of UV radiation, whereas shocks produced by jets can lead to ionization of gas in both high-mass and low-mass YSOs \citep{anglada18}.


\begin{table*}[t]
\caption{Parameters of VLA observations (Project ID: 22A-086).}
\label{table:1}
\centering
\begin{tabular}{c c c c c c c}
\hline\hline
& & & \multicolumn{2}{c}{Clean beam } &\multicolumn{2}{c}{$\sigma$ }\\
\cline{4-7}
Date    &Starting time  &Ending time  & Size  & P.A.  & I  & V  \\
(dd-mm-yy)  & (UTC) & (UTC) & ($''\times''$) & (deg.) & ($\mu$Jy~beam$^{-1}$) & ($\mu$Jy~beam$^{-1}$) \\
\hline
   05-03-22            & 10:47:47 & 12:17:32 & $0.36\times0.15$          & 168             & 3.2                 & 3.1                 \\    
   09-03-22            &  12:06:19 & 13:35:59 & $0.30\times0.14$          & 179              & 2.9                 & 2.9                 \\
   10-03-22a  &  10:28:08 & 11:57:54 & $0.36\times0.14$          & 168             & 3.2                 & 3.2                 \\
   10-03-22b  & 11:57:57  & 13:27:39 & $0.31\times0.14$          & 177              & 3.1                 & 3.2                 \\
   26-03-22            &  12:51:58 & 14:21:43 & $0.36\times0.13$          & 21              & 3.2                 & 3.2                 \\ 
   27-03-22            & 10:31:02  & 12:00:43 & $0.32\times0.14$          & 174              & 3.1                 & 3.2                 \\ 
   30-03-22            & 09:16:35  & 10:46:16 & $0.40\times0.15$          & 159             & 4.4                 & 4.5                 \\ 
   31-03-22            & 09:43:49  & 11:13:32 & $0.35 \times 0.15$  & 162             & 3.2                 & 3.2                 \\ 
\hline
  All epochs combined   & & & $0.31\times0.15$ & 170 & 1.2  & 1.1\\
\hline    
\end{tabular}
\tablefoot{The first three columns indicate the date and time for which each observation was performed. The clean beam sizes and position angle (P.A., the eastward angle between the North and the major axis) have been calculated from measurements of the major and minor axis full-width half-maximum (FWHM), using the CASA task {\tt imfit}. The noise $\sigma$ for Stokes I and V has been calculated from general {\tt casaviewer} statistics by defining a region away from the source in the Stokes I and V maps, for each epoch.}
\end{table*}

Besides thermal emission,  YSOs emit at radio frequencies also via non-thermal processes. A recent study detected such emission in 50\% of a sample of 15 high-mass YSOs, with clear non-thermal lobes in 40\% of them \citep{Non_thermal_radio_emission_from_jets}. Another survey in the Perseus star-forming region revealed that nearly 60\% of YSOs in the region exhibit radio properties indicative of a non-thermal origin \citep{Perseus_region}. For instance, a recent VLA study concluded that the emission in G14.2-N is mainly dominated by non-thermal sources \citep{Diaz}. The non-thermal radio emission is generally attributed to processes involving magnetic fields \citep{Dulk}. The synchrotron/gyrosynchrotron/gyroresonance radiation caused by the gyration of ultra/mildly/non-relativistic particles is incoherent, has a slightly negative spectral slope, and can present a modest circular polarization (if particles are not ultra-relativistic). This kind of emission has been observed in shocks in jets \citep{Non_thermal_radio_emission_from_jets,anglada18}, accreting disks and YSO photospheres \citep{Anglada}, in addition to HII regions \citep{HII_regions}. 
Additionally, plasma emission and electron cyclotron maser are two coherent mechanisms which give rise to coherent, highly circular polarized emission, which is seen in YSOs (e.g., T-Tauri star \citealt{T_Tauri}), cool dwarfs \citep{Yiu,pineda23} and planets \citep{zarka07,Zarka1999,Stevans}.

In this article, we present a comprehensive study of radio emission at 15 GHz, from [BHB2007]-01, a K7-type YSO located in Barnard 59, which is the only active star formation site in the otherwise quiescent Pipe nebula at a distance of $163\pm5$ pc \citep{Distance_BHB07} from Earth. This system is probably less than 1 Myr old \citep{Covey,Alves_2020} and comprises a low-mass young star surrounded by a 107 au-wide, almost edge-on ($i \simeq 75^\circ$) disk with a vast 70 au-wide gap, potentially harboring a very young sub-stellar object \citep{Alves_2020,Zurlo}. The YSO has been reported to have Br$\gamma$ and Pa$\beta$ emission lines in the near-infrared, indicating ongoing accretion \citep{Covey}, while the observed gap is devoid of large dust particles, but probably full of molecular gas and small dust particles. Observations at 226 GHz with Atacama Large Millimeter/submillimeter Array (ALMA) revealed a warm (100 K) and compact molecular gas component (traced by CO) near a cm-wave radio emission seen at 22.2 GHz with the VLA in the south-eastern part of the disk gap \citep{Alves_2020}. This pointed at the presence of a very young, possibly still accreting, sub-stellar object in the dusty gap. A subsequent near-IR study conducted by the Very Large Telescope/NACO revealed a point source at a distance of 50 au \citep{Zurlo} in L'-band, coinciding with the position of the faint emission seen in radio by the VLA. This marked the first potential radio detection of a sub-stellar companion of a very young star. 
Using evolutionary models \citep{Zurlo}, the mass of
the point source was constrained to be between 37–47 $M_J$, while it can be smaller if the object is surrounded by a circumplanetary disk. This was consistent with 1D
hydrodynamical modeling \citep{Alves_2020} which needs an object of 4–70 $M_{J}$ to carve out a 70 au gap in the disk.

Here we analyze new VLA data in Ku band (15 GHz), in order to shed light on the system. We perform a detailed analysis on 8 observations, focusing on the time variability and polarization properties from the central star, and looking for the emission from the sub-stellar object. The paper is organized as follows: Sect. \ref{sec:observations} provides a detailed account of the observations and the associated data reduction strategy. In Sect. \ref{sec:results}, we present the results, segmented into distinct subsections focusing on time variability of the observed signal, circularly polarized emission, the spectral energy distribution, and the sub-stellar object. Finally, we culminate with the discussion in Sect. \ref{sec:discussion}.


\section{Observations and data reduction} \label{sec:observations}

The target, [BHB2007]-1 ($\alpha_{J2000} = 17^{\rm h}11^{\rm m}03\fs 92$, $\delta_{J2000}=-27^{\circ}22'55\farcs47$)
had 8 dedicated VLA observations in March 2022, each lasting 1.5 hr (12 hr in total, VLA/22A-086). 
Table \ref{table:1} summarizes the observations, indicating the date, size and orientation of the synthesized beam, along with the background root-mean-square value (rms) $\sigma$ in Stokes I and V. As the nature and origin of radio emission from our target is an open question, we chose Ku band (15~GHz, covering a frequency range between 12 and 18~GHz) to detect emission by various possible mechanisms, both thermal and non-thermal. Continuum standard mode was chosen to allow for maximum bandwidth coverage and 
the observation made use of the telescope's most extended configuration, A. The combination of the frequency band and the telescope configuration accounted for an angular resolution of $\sim$ 0.20". 
The spectral set-up consisted of 48 spectral windows, each having a bandwidth of 128 MHz and composed of 64, 2 MHz-broad, channels.

The first observation (March 5$^{\rm th}$) potentially suffered from some issues on the pointing (priv. comm.), so we analyzed this particular dataset with extra care, looking for hints of unreliable results. 
We find that the images and statistics are consistent with the properties of those for the other days; this, together with the absence of warning issues in the typical diagnostics, suggests that the data quality is not negatively impacted.

\subsection{Calibration and Imaging}

During each observation, J1700$-$2610 was used for gain calibration. Additionally, 3C286 and J1650$-$2943 were observed as absolute flux and band-pass calibrators, respectively, once per observing session. The on-source time was $\sim$ 0.91 hrs in each epoch.


The entire dataset underwent processing via CASA  \citep[Common Astronomy Software Applications;][]{CASA}. The calibration of the data adheres to the standard CASA VLA pipeline. The maps have been constructed with the CASA {\tt tclean} task by using the {\tt standard} gridding algorithm and employing a {\tt natural}
weighting scheme across all image planes to optimize sensitivity, while CARTA \citep[Cube Analysis and Rendering Tool for Astronomy; ][]{carta_1,carta_2} was used for visualizing the maps. During the deconvolution process, we used the multi-frequency synthesis and fit the frequency dependence of the observed emission with a Taylor series expansion with {\tt nterms=2}, to account for the spectral index of the emission.

By stacking together the target scans from the 8 observations, we reached a background rms of $\sigma_I= 1.2$ $\mu$Jy~beam$^{-1}$  in Stokes I and $\sigma_V=1.1$ $\mu$Jy~beam$^{-1}$ in Stokes V. 
Table \ref{table:1} shows the characteristics of the resultant clean beam for each observation, along with the position angle (P.A.), and the corresponding $\sigma$ values for Stokes I and Stokes V.

\begin{table*}
\caption{Statistics of fluxes and deconvolved emission, from Gaussian fitting of the clean images of each day.} 
\label{table:2}      
\centering          
\begin{tabular}{c c  c c c c c c }     
\hline\hline       
& \multicolumn{3}{c}{Deconvolved component(Stokes I)} &  \multicolumn{2}{c}{Flux (Stokes I) } & \multicolumn{2}{c}{Flux (Stokes V) } \\    
\cline{2-8}
Date & Major axis & Minor axis & P.A. & Peak & Integrated     & Peak         & Integrated\\   
(dd-mm-yy) & (mas)& (mas) & (deg.) & ($\mu$Jy~beam$^{-1}$) & ($\mu$Jy) & ($\mu$Jy~beam$^{-1}$) & ($\mu$Jy) \\
\hline                    
05-03-22  & $192\pm26$  & $60\pm6$   & $156\pm6$  & $177\pm3$ & $216.3\pm7.7$  & $29.2\pm2.1$ & $25.4\pm4.0$ \\  
09-03-22  & $118\pm33$  & $82\pm15$  & $176\pm36$ & $146\pm4$ & $181.4\pm7.9$  & $<8.7$       & \textemdash\\
10-03-22a & $174\pm33$  & $80\pm14$  & $178\pm27$ & $147\pm3$ & $172.1\pm6.9$  & $<9.6$       & \textemdash\\
10-03-22b & $189\pm28$  & $85\pm3$   & $164\pm8$  & $135\pm2$ & $158.3\pm5.1$  & $<9.6$       & \textemdash\\
26-03-22  & $151\pm40$  & $99\pm9$   & $21\pm8$   & $114\pm4$ & $173.4\pm9.0$  & $19.9\pm2.9$ & $37.4\pm8.2$ \\
27-03-22  & $231\pm37$  & $78\pm16$  & $164\pm4$  & $88\pm3$  & $128.1\pm7.8$  & $<9.6$       & \textemdash\\
30-03-22  & $255\pm100$ & $101\pm34$ & $155\pm9$  & $75\pm5$  & $108.0\pm11.0$ & $<13.5$      & \textemdash\\
31-03-22  & $192\pm30$  & $83\pm11$  & $156\pm6$  & $101\pm2$ & $132.7\pm5.3$  & $<9.6$       & \textemdash \\
\hline   
   All days combined   & $147\pm15$   & $69\pm8$   & $154\pm7$  & $127\pm2$ & $157.5\pm3.5$  & $9.4\pm0.9$     & $8.1\pm1.7$ \\ 
\hline                  
\end{tabular}
{\tablefoot{The major and minor axis (FWHM) report the image component size (deconvolved from the beam using {\tt gaussfit}) and have been taken from the output of CASA task {\tt imfit} for Stokes I images, along with the value of the position angle. Further, the peak and integrated flux, and the related errors, for both the polarisations have also been calculated by {\tt imxfit}, by performing a Gaussian fitting. All errors are 1$\sigma$, and include only the uncertainty on the fit, and not the overall rms estimated from the images and quoted in Table \ref{table:1}. For the days without significant Stokes V signal, we have provided 3$\sigma$ upper limits for the peak flux. In the two Stokes V images with a significant detection, {\tt gaussfit} provided a size of the component $(423\pm126)$ mas $\times \, (150\pm32)$ mas for the March 26$^{\rm th}$ observation, while it could not deconvolve the component from the beam for the March 5$^{\rm th}$ observation, hence a size estimation could not be provided.}
}
\end{table*}

\subsection{Flux-extraction, time-variability and further reduction}

We employed the CASA task {\tt imfit} to conduct measurements on the flux densities, sizes, and positions of the Stokes I and Stokes V components of the emission.
We also used the {\tt maxfit} function in CASA to derive the peak flux intensity values, which were in accordance with the peak values obtained from 2D-Gaussian fitting.

Further, to investigate the time variability in the signal in each session, we used the CASA task {\tt visstat} that calculates the flux statistics in the visibility plane. We looked for potential variability in the real part of the amplitude by splitting the data corresponding to our target field over small time intervals of 10 s and 30 s, and integrating it over all the spectral windows. Following this, the possible periodicity in the flux was determined by applying the Generalized Lomb-Scargle (GLS) periodogram, which is a commonly used statistical tool designed to detect periodic signals in unevenly spaced observations \citep{GLS}. It must be noted that there were no major bright sources in the target's field of view to potentially interfere with the flux measurements of the target. Another segment of the analysis also focused on studying the variation of flux with frequency.


Since the young stellar object ([BHB2007]-1) is radio-brighter than the plausible sub-stellar companion, the emission from the former can potentially overpower the emission from the latter to significant levels, thereby making the latter undetectable in the maps. Considering this, we also produced the images of the target field after subtracting the emission from the YSO with the CASA task {\tt uvsub}. To do so, a clean model of the YSO emission was made for each observation and then subtracted from the individual visibility sets.

\section{Results} \label{sec:results}

\subsection{Combined image}

\begin{figure}
\centering
\includegraphics[width=\hsize]{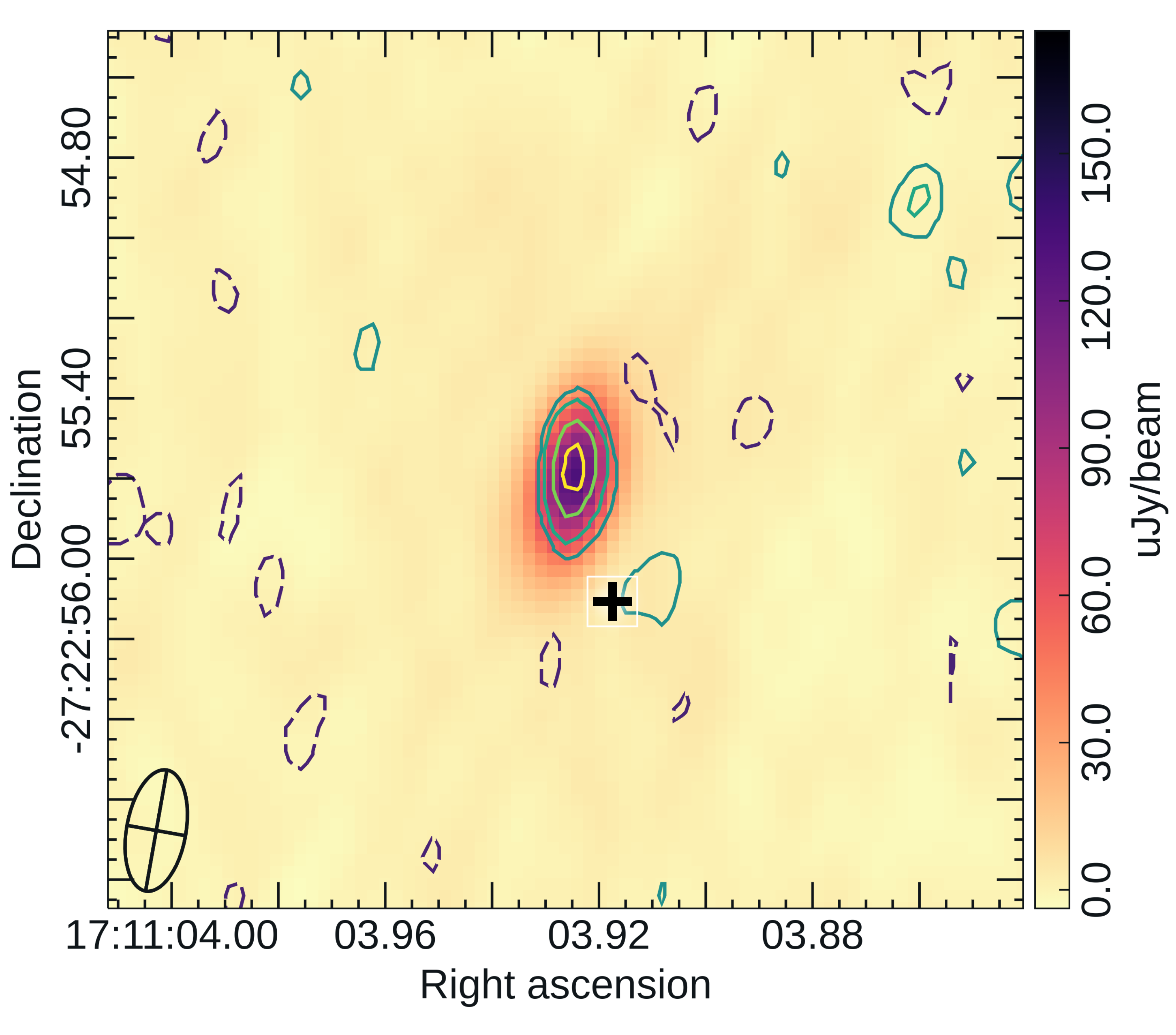}
\caption{Continuum map, centered at the position of the YSO, combining the eight observations of this work. The color scale indicates the Stokes I emission. Contours identify the Stokes V emission, at a level of -3, -2, 2, 3, 5, and 7 times the Stokes V rms, which is $\sigma_V\sim 1.1 \mu$Jy~beam$^{-1}$ . Negative contours are shown in purple dashes, and positive ones in green-yellow solid lines. The cross (\textbf{+}) marks the position of the tentative brown draft seen in previous VLA observations. The synthesized beam is shown in the lower left corner.}
\label{combined_image}
\end{figure}

 Figure ~\ref{combined_image} shows the cleaned image produced by combining all observing days and all channels (Fig. \ref{combined_image}). We have obtained a clear detection of the YSO, with a peak intensity of $\sim$127~$\mu$Jy~beam$^{-1}$ in Stokes I, corresponding to a signal-to-noise ratio S/N$\sim 100$. The YSO has also been detected in Stokes V with a peak flux of $\sim$9.4~$\mu$Jy~beam$^{-1}$, indicating an average polarization fraction of $\sim7.5\%$. The $\sim 20\%$ disparity in the peak versus integrated fluxes and the sizes of the deconvolved components, which are just larger than the clean beam (see Table \ref{table:2}), suggest that the emission might be slightly resolved.

There is no conclusive indication of emission from the tentative position of the brown dwarf (marked by a cross in Fig.\ref{combined_image}). The expected position has been calculated considering the previous radio observation by \cite{Alves_2020} on 2016, October 15$^{\rm th}$ (VLA 16B/260), from which we adopted the coordinates for the tentative sub-stellar object ( $\alpha_{J2000} = 17^{\rm h}11^{\rm m}03\fs 91$, $\delta_{J2000}=-27^{\circ}22'55\farcs79$1) and applied the correction considering the relative shift in the position of YSO compared to the previous VLA detection (\citep{Alves_2020}). The shift in the position of the YSO is within the proper motion reported for Gaia DR3 \citep{Gaia_mission,Gaia_dr3_summary}, source ID 4108624199978985984 (adopted by \citealt{Zurlo} as well): $\mu_{\alpha} = -3.67\pm0.60$~mas~yr$^{-1}$, $\mu_{\delta} = -17.94\pm 0.41$~mas~yr$^{-1}$. 
Also, we can estimate the orbital period of the putative brown dwarf to be $\sim 2.3$ centuries, assuming Keplerian circular orbit at $\sim 50$ au separation and stellar mass $M\sim 2.2 M_\odot$ \citep{Alves_2020}. The corresponding orbital motion covered in $\sim 5.5$ years would be $\sim 2.4 \%$ of the orbit (meaning $\sim 40$ mas), comparable with the proper motion of the system. We account for an overall $\sim 40$ mas uncertainty in the position of the sub-stellar object, which is indicated by the cross size.

The $2\sigma$ contour in Stokes V in fig. \ref{combined_image}, almost coincident with such location, is suggestive, but statistically not sufficient to claim a detection of the sub-stellar object. Considering the rms for the total integrated map, we set a $3\sigma$ upper limit of $\sim$3.6~$\mu$Jy to the 15~GHz emission from the putative sub-stellar object.

\begin{figure*}
\centering
\includegraphics[width=1\textwidth]{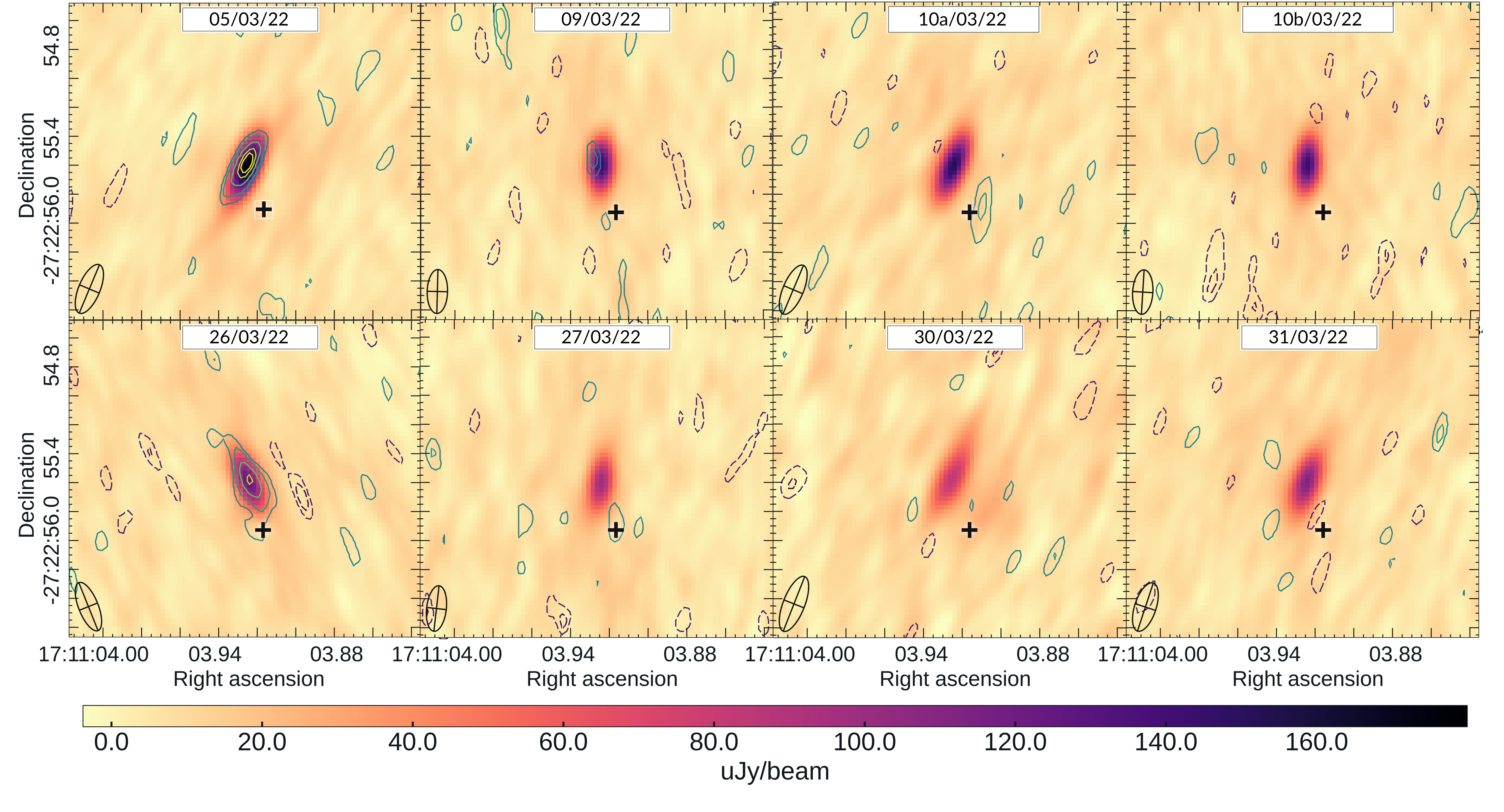}
\caption{Images for each observation (day indicated in the label) for Stokes I, with Stokes V overlaid in contours. The contour levels are color coded and represent -3,-2, 2, 3, 5, 7, and 8 times the $\sigma$, which ranges from $3-4\mu$Jy~beam$^{-1}$ (see Table~\ref{table:1}). The synthesized beam for each day is shown in the lower left corner, while the cross \textbf{+} marks the position of the potential sub-stellar companion as seen in previous observations. The position has been corrected, taking into consideration the proper motion of the YSO from the Gaia database.}
\label{fig:images_days}
\end{figure*}

\subsection{Variable flux in Stokes I across the different epochs}

As the observations were spread over a month, we also searched for possible variability in the signal from the YSO over different observing days, and within each of them. A detailed description is presented in the following sections. We report in Table \ref{table:2} the detailed analysis for each epoch and for the combined image discussed above. We show the images of Stokes I (in color) and Stokes V (in contours) for each observing run in Fig.\,\ref{fig:images_days}. The YSO is clearly detected in all of them in Stokes I. A right circular polarized signal is clearly detected on March 5$^{\rm th}$ and March 26$^{\rm th}$, with a marginal detection on March 9$^{\rm th}$, and no detection in the other five observations.

In correspondence with the position of the sub-stellar object (marked by a cross with the width corresponding to the aforementioned uncertainty), we only see a marginal, not significant detection during the first observation of March 10$^{\rm th}$ and March 27$^{\rm th}$ in Stokes V (S/N $\sim 2-3 \sigma$). In Stokes I, it is much more difficult to evaluate excess at few tens $\mu$Jy level at the same position because of the brighter central emission (we have also tried to use a {\tt briggs} weighting to enhance the resolution): visually, the only suggestive case is on March 30$^{\rm th}$ with a S/N $\sim 5 \sigma$ (possibly helped by the lowest flux from the central star on that day). We can neither claim or discard an excess at the possible sub-stellar position.

\begin{figure}
\centering
\includegraphics[width=\hsize]{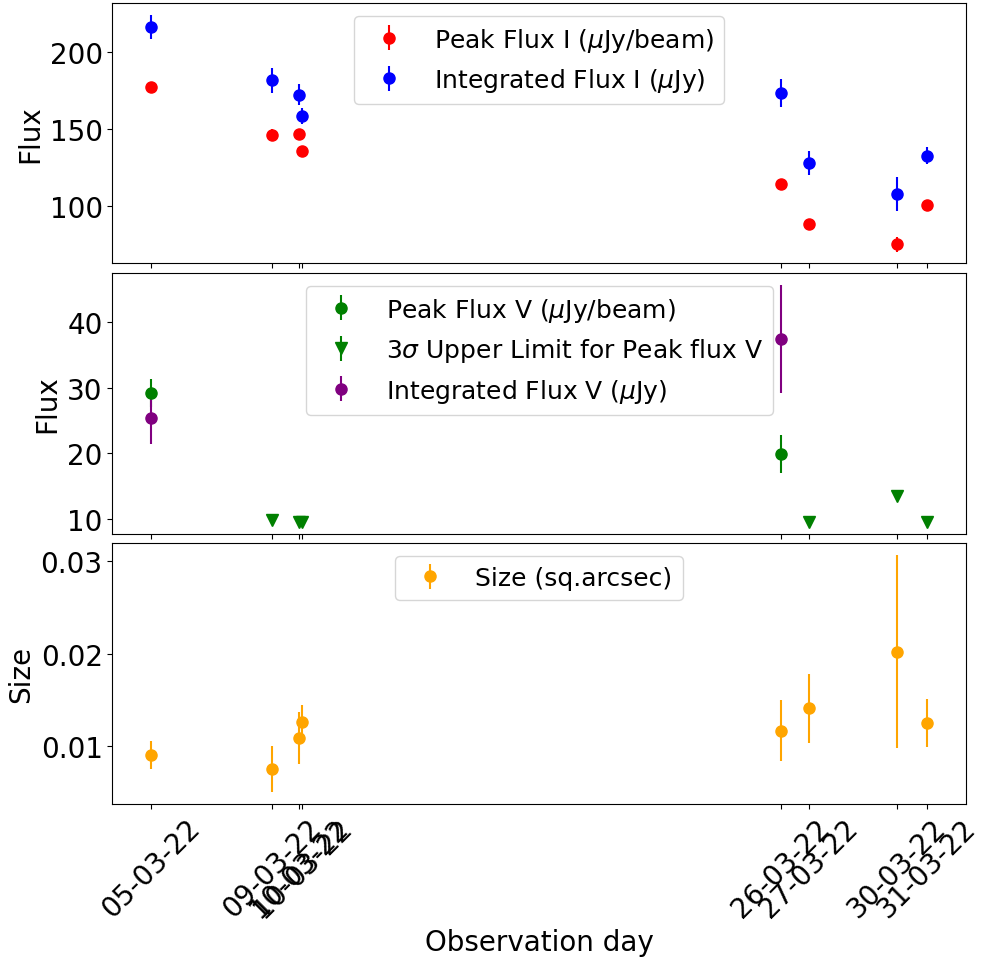}
\caption{Variability over the   over the eight observational epochs. We show the peak flux intensity and the integrated flux density in Stokes I (top) and Stokes V (middle panel), and the size of the deconvolved component of the emission in Stokes I, calculated by using the values and errors of the major and minor axes in Table \ref{table:2}. Vertical bars show 1$\sigma$ errors. For the days without a Stokes V detection, we have provided $3\sigma$ upper limits for the peak flux.}
\label{flux_days}
\end{figure}

Fig.~\ref{flux_days} shows the total flux density in Stokes I, with significant variability across the observational window spanning 27 days. During these eight epochs, the flux values fluctuated, from the highest on the first day, March 5$^{\rm th}$, with an integrated flux of $ 216~\mu$Jy, to the minimum value of $\sim 108~\mu$Jy on March 30$^{\rm th}$. Such variations are up to a factor $\sim$2, and much higher than the flux errors. The variable behavior is seen in both peak and integrated fluxes. Overall, the variations did not exhibit any discernible or predictable pattern. As a matter of fact, between the observations of March 26$^{\rm th}$ and 27$^{\rm th}$, the integrated flux decreased by $\sim 1/4$ in 22 hours, pointing to substantial variations in $\lesssim$ 1 day. Similarly, there is a significant flux increase between March 30$^{\rm th}$ and 31$^{\rm st}$. On the other hand, the three closest observations between March $9^{\rm th}$ and $10^{\rm th}$ show integrated fluxes compatible within 2$\sigma$ to each other. These erratic, $\sim$ day variability timescales can be indicative of processes coming from the YSO itself or the immediate surroundings. 

The signs for a slightly resolved extended emission, discussed above for the combined images, are consistently seen in each observation, too: the peak flux is systematically $\sim 20-30\%$ lower than the integrated flux, and the reconstructed values of the deconvolved Stokes I components are slightly larger than the clean beams (see Table \ref{table:2}). A complementary analysis of the amplitude versus uv-distance profiles did not shed further light in this aspect, because the fluctuations are not small enough to distinguish between a point source and slightly resolved emission. 

\begin{figure}
\centering
\includegraphics[width=\hsize]{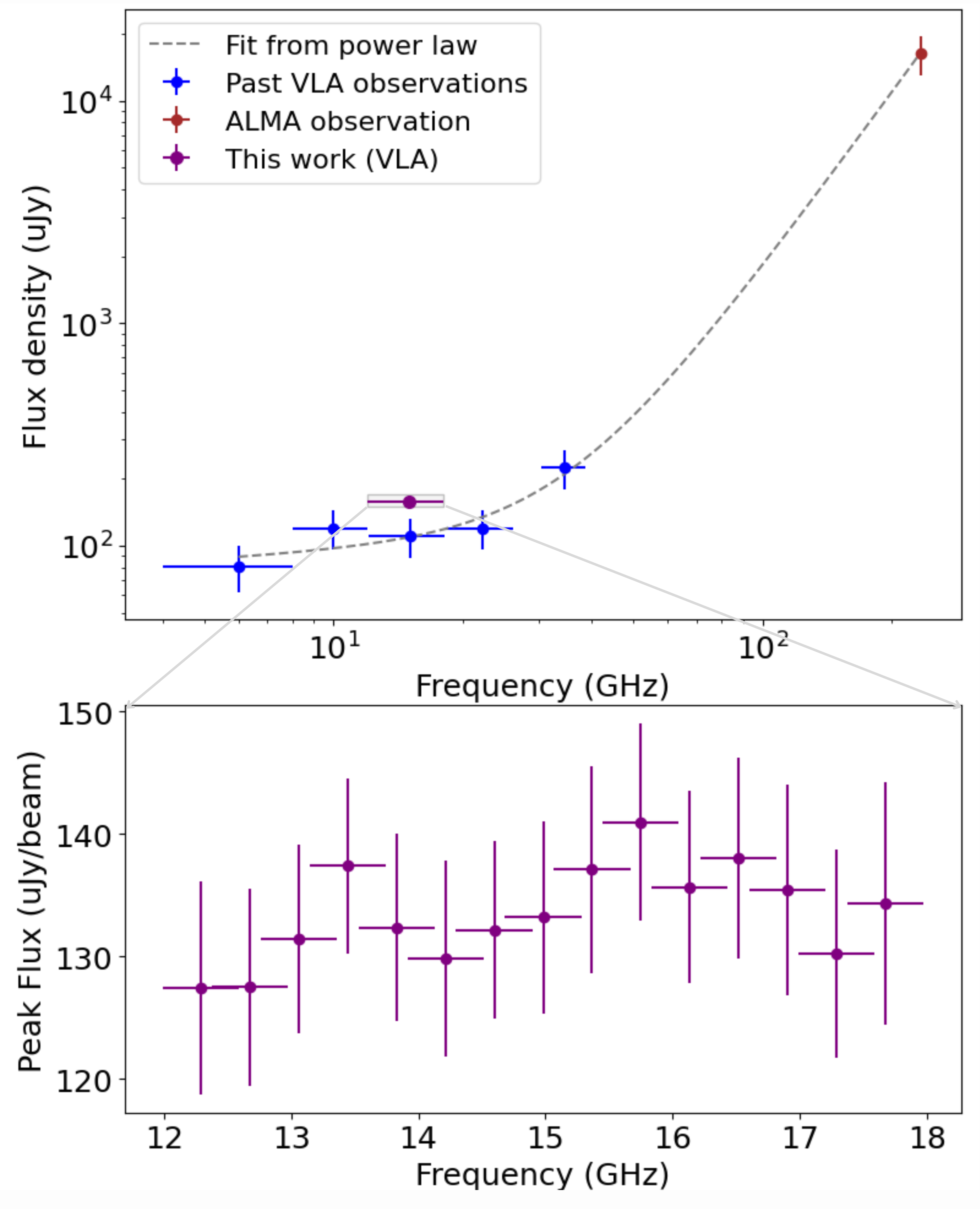}
\caption{{Top: A broadband spectrum of flux density for the YSO in Stokes I, considering all available VLA (4-35 GHz) and ALMA (230 GHz) data \citep{Dzib_2016, Alves_2020, Zurlo}. The horizontal bars represent the bandwidth for each observed band. Bottom: Spectral distribution of mean flux in Stokes I for the current VLA observations averaged over all days and over 3 spectral windows, ie. 384 MHz. The vertical error bars correspond to the 1$\sigma$, while the width of the horizontal bars is equivalent to the bandwidth of 384 MHz over which each data point has been averaged.}}
\label{Fig:SpecIndex_VLA-ALMA}
\end{figure}



\begin{figure*}[ht!]
\centering
\begin{minipage}[b]{0.49\textwidth}
\includegraphics[width=\linewidth]{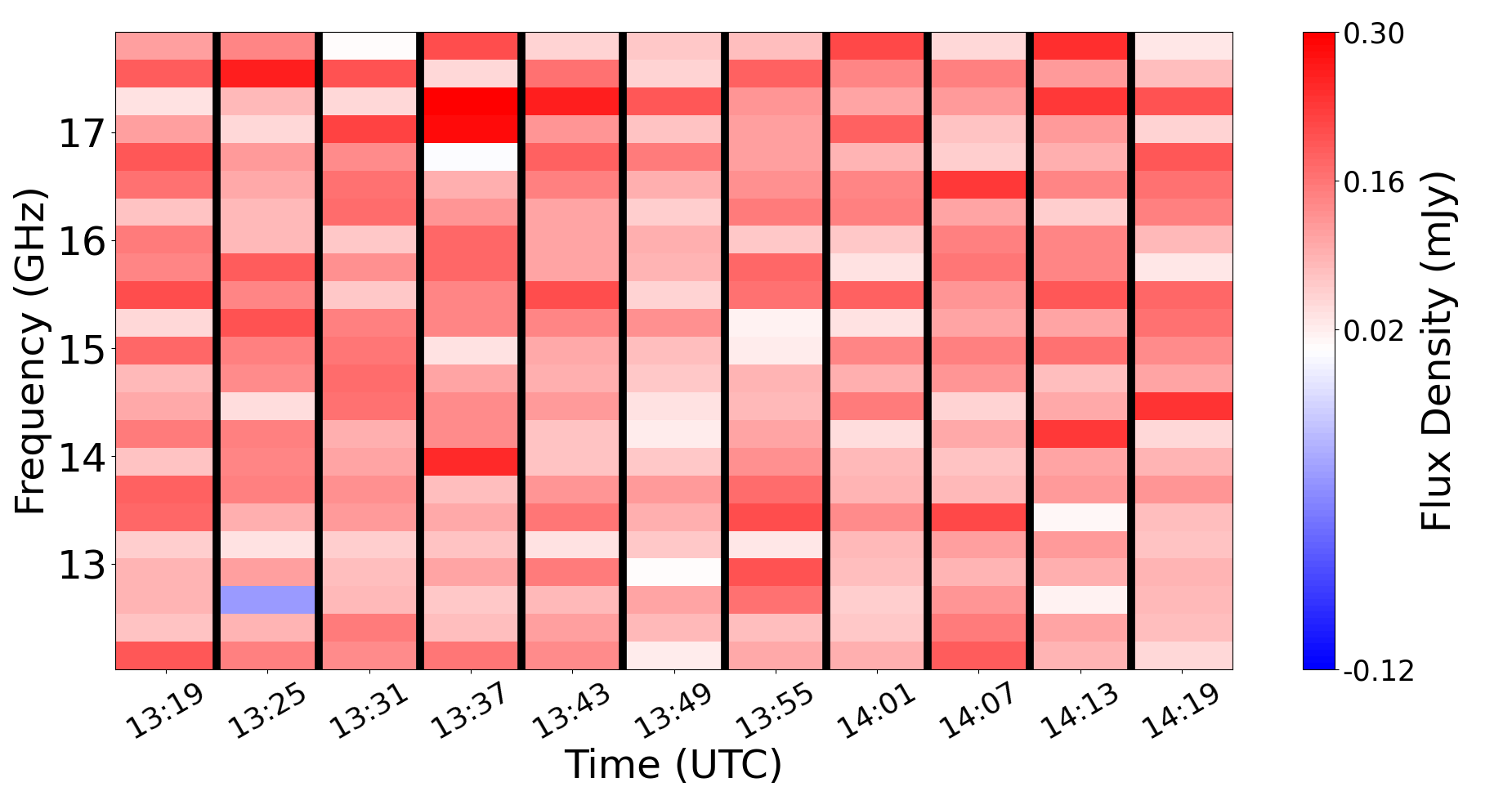}
\end{minipage}
\hfill 
\begin{minipage}[b]{0.49\textwidth}
\includegraphics[width=\linewidth]{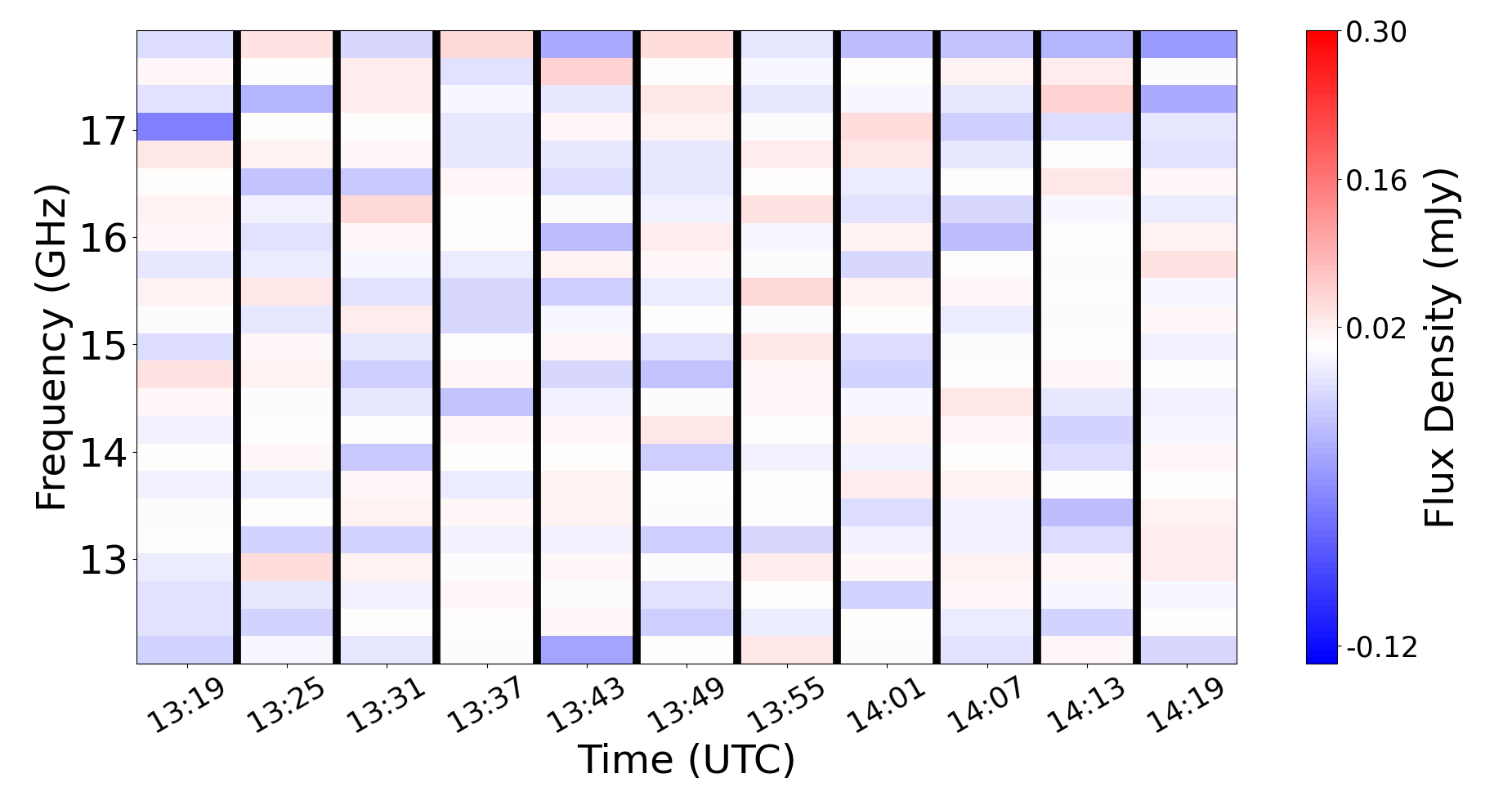}
\end{minipage}

\vspace{10pt} 
\begin{minipage}[b]{0.49\textwidth}
\includegraphics[width=\linewidth]{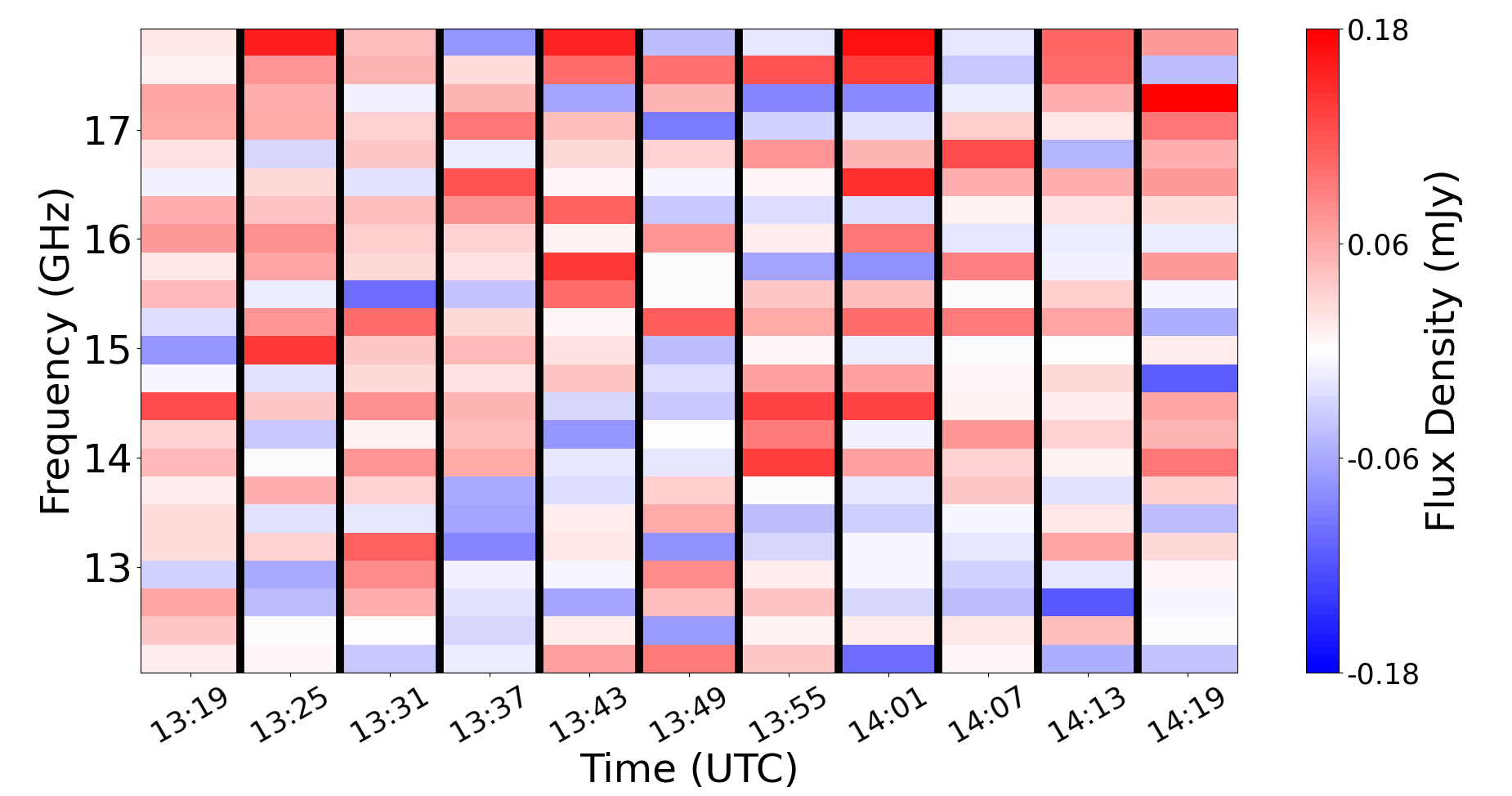}
\end{minipage}
\hfill 
\begin{minipage}[b]{0.49\textwidth}
\includegraphics[width=\linewidth]{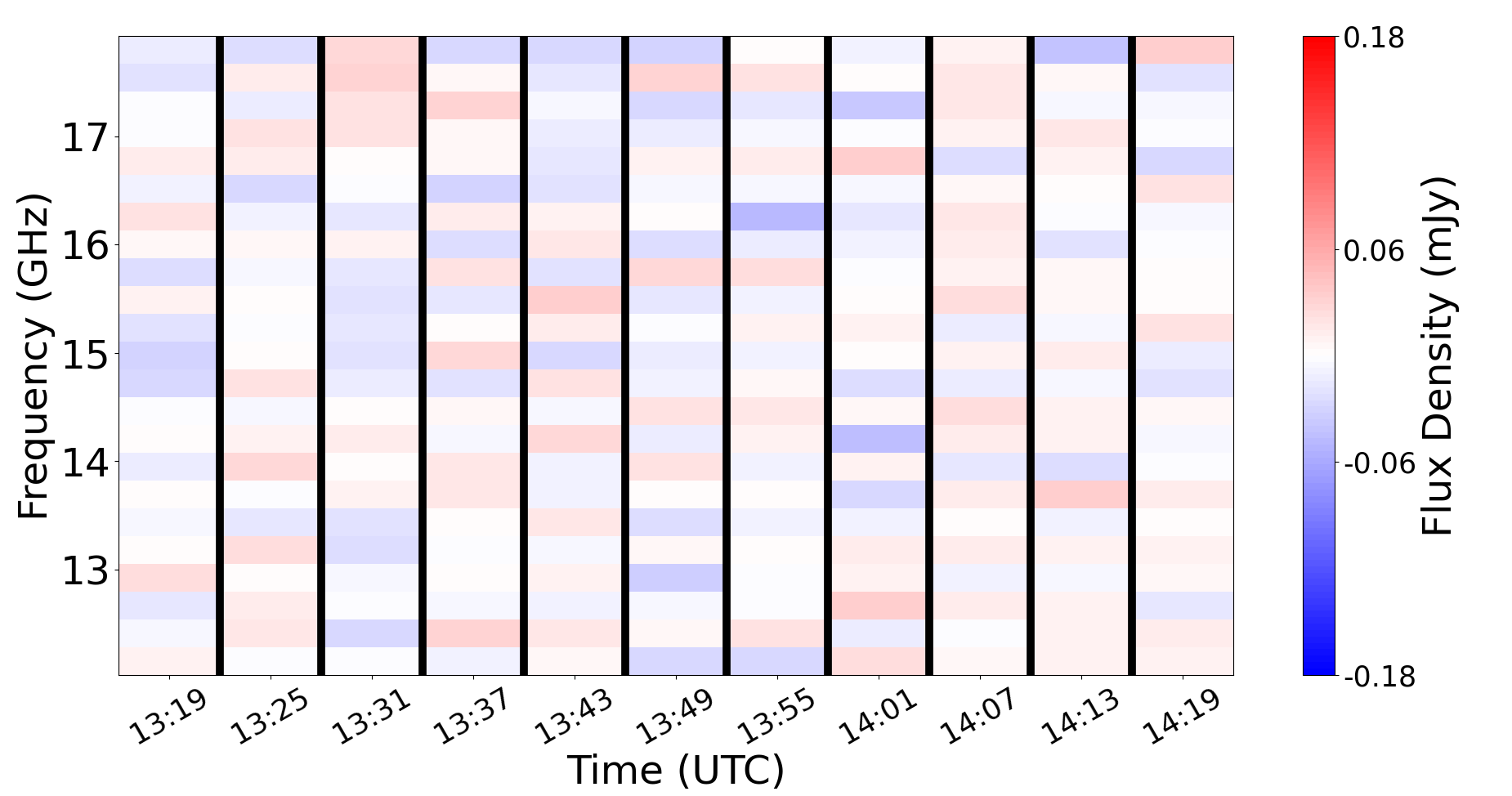}
\end{minipage}

\caption{Dynamic spectrum of a region encompassing the source (left) and another region away from the phase center for the March 26$^{\rm th}$ observation, in Stokes I (top) and Stokes V (bottom). The dynamic spectrum has been obtained selecting the source from the image plane and integrating over 256 MHz-wide spectral windows, for each scan (approximately 6 minutes long). On average, the rms in each pixel is $\sim$70~$\mu$Jy~beam$^{-1}$ for each scan in both Stokes I and Stokes V. The thin vertical black lines between target scans represent the time spent on the phase calibrator ($\sim$48~s). The panels on the right side have been plotted only to present a visual comparison between the flux values during the presence and absence of a radio source. To obtain these spectra, we used the CASA task {\tt specflux} and {\tt imstat} after generating the cube images with {\tt tclean}, for every source scan.}
\label{fig:dynamic_spectra}
\end{figure*}


\subsection{Circularly polarized signal in two epochs}

In order to look for a circularly polarized signal, we analyzed the Stokes V images as well. We got significant Stokes V detection with $\sim 9 \sigma$ and $\sim 6 \sigma$ in the March 5$^{\rm th}$ and March 26$^{\rm th}$ observations, with a polarization fraction of about $16-18 \%$. For the rest of the days without a detection, we set 3$\sigma$ upper limits of the order of $10~ \mu$Jy/beam, corresponding to circular polarization upper limits in the range $\sim 5-12\%$. We also combined the remaining six datasets without a Stokes V detection, in order to slightly improve their rms, but only found a non significant $S/N\sim 2\sigma$ Stokes V signal.

Table~\ref{table:2} also shows the peak and integrated flux densities of the Stokes V signal (or the 3$\sigma$ upper limit in the absence of detection) for each epoch. The sporadicity of the emission seen here is typical of circularly polarized radiation driven by plasma emission or Electron Cyclotron Maser (ECM) emission. High polarization fractions  are expected for these mechanisms \citep{Dulk}, which can be consistent to the moderate values observed here if one considers the superposition with a persistent, unpolarized component. In this sense, it is suggestive to note that the difference of Stokes I flux between the days (5$^{\rm th}$, 26$^{\rm th}$) with Stokes V detection and the following available observations ($9^{\rm th}$, $27^{\rm th}$) are positive and of the same order of the detected Stokes V. We do not have enough observational cadence to confirm this suggestion, that would point to a $\sim 100\%$ polarized component in addition to the persistent one.

Alternatively, incoherent gyro-synchrotron emission from mildly relativistic electrons can also provide modest circular polarization. We also note that, considering a tentative $3\sigma$ detection during the March 9$^{\rm th}$ observation and the upper limits in the remaining observations, we cannot discard a persistent but variable weak polarization, which would again favour gyro-synchrotron.



\subsection{Frequency dependence of the emission}

We have investigated the frequency dependence of the emission, by combining all days, using the CASA task {\tt specflux}. As shown in the lower panel of Fig. \ref{Fig:SpecIndex_VLA-ALMA}, the flux-frequency curve looks flat within the uncertainties, with the flux density only slightly increasing with frequencies. Additionally, we inferred the spectral index on a scan-by-scan basis, inferred by looking at the slope between 12--15 GHz and 15--18 GHz flux density. 
All spectral indexes are close to flat or slightly positive ($-0.16<\alpha<0.5$) and compatible with each other within the (large) errors.

In order to put these results in the context of a broader frequency range, Fig.~\ref{Fig:SpecIndex_VLA-ALMA}
also shows the global spectrum where past VLA and ALMA observations are considered, from 4 to 230 GHz \citep{Dzib_2016, Alves_2020}. Overall, there is an apparent change of slope happening just around the 15~GHz band studied here. Indeed, this is consistent with the almost flat spectrum seen in our data. The change of slope is probably due to the co-existence of the (gyro-)synchrotron emission (negative slope, dominating at low frequencies) and free-free emission (positive slope, higher frequencies). Note also that the flux detected in the current observations is slightly higher than in the past ones.

\subsection{Short-term variability}

We investigated the possible contributions to the flux variability from short flaring events. We examined fluctuations in the peak flux intensity during individual observing sessions. 

First, we have looked at the dynamic spectra, shown in Fig. \ref{fig:dynamic_spectra} for March 26$^{\rm th}$, one of the two days with a clear Stokes V detection. We show the results for Stokes I (top panel) and Stokes V (bottom). It is possible that maybe, due to the rms fluctuations, there is no visible spike (in time) or spectral features for both Stokes I and V. The emission seems to increase slightly with frequency, in agreement with the slightly positive spectral index found in that day's observation when split into two bands as mentioned before ($\alpha = 0.33 \pm 0.25$).

To investigate on shorter timescales, we divided the data from each session into shorter integration times of 10, 30 and 60 seconds, by integrating the data over all the channels to have enough S/N. For this purpose, we used the CASA task {\tt visstat} to calculate the statistics separately for the LL and RR correlators and obtained Stokes I ((RR+LL)/2) and Stokes V ((RR-LL)/2) statistics.
We did not find any clear, structured spikes in any epoch, although there were a few single-point spikes with S/N$>5$, but with no conclusive periodic behavior. (See Appendix A for the plots showing the time series for an integration time of 60 s for all the days).
In order to look for periodicity in these data sets, we also obtained the GLS periodogram \citep{GLS}, which gave us no statistically significant peak (False Alarm Probability <1$\%$) in the Fourier space at any epoch.

In summary, we did not find any sign of short (minutes or sub-minutes) clear variability or spikes.

\section{Discussion and Conclusions} \label{sec:discussion}

YSO [BHB2007]-1 is a peculiar object, showing interesting properties across the radio/mm spectrum. Here we have shown a detailed analysis of eight observations taken with VLA at 15 GHz. We clearly detect the radio emission from the central YSO, with a slightly resolved emission. The observed radio emission also has a mild variability on $\sim$ days timescales, with no clear minute/sub-minute variability trend. 

The spectral index is slightly positive, not totally incompatible with being flat. Combined with previous data from VLA and ALMA between 10 and 230 GHz, there seems indeed to be a spectral turning point right at the Ku band (15 GHz), that is being explored in this study. This is compatible with the presence of two components: a free-free emission component and a non-thermal 
component. The free-free emission is probably related with collimated winds observed in CO \citep{Alves_2020} and tentatively detected in the near-IR \citep{Zurlo}, while the non-thermal component, predominantly (gyro-)synchrotron, has a negative slope and dominates at lower frequencies. Therefore, the detection of a non-thermal component becomes more difficult at higher frequencies where free free emission becomes prominent \citep{Gudel_1}. 

In agreement with this interpretation, we found two epochs with a clear Stokes V detection of $\sim 16-18\%$. The lack of such detection in other epochs is compatible with the presence of a sporadic, possibly moderate or highly circularly polarized non-thermal component, coming on top of the persistent, thermal one. Among non-thermal processes, gyro-synchrotron emission can provide moderate values of circular polarization \citep{Dulk, Gyrosynchrotron_andre}, while the coherent plasma emission or ECM, have larger polarization fractions \citep{Dulk, ECM, UV_Cet,ULD, Yiu,Callingham}. 

The Stokes V emission seems to be slightly resolved as well. In this scenario, the non-thermal component would originate from electrons in the surroundings which gyrate around the magnetic fields. The resulting polarization fraction strongly depends on the electron energy, and can have moderate values for mildly relativistic electrons, with Lorentz factors $\gamma \leq 2-3$ \citep{Dulk}. In this context, the jet-induced shocks could also be accelerating the electrons to such mildly relativistic energies, thereby contributing to the observed non-thermal emission \citep{High_energy_processes,R_Coronae, Ainsworth_Jets}. 
It is worth noting that in the epochs where we have clearly detected circularly polarized emission, the integrated flux density for stokes I emission appears to have increased by a similar amount (within the errors) of the stokes V, with respect to the observations made just after the stokes V detection (See Table \ref{table:2}). Since the observed level of circular polarization depends also on the contemporaneous presence of the unpolarized components, in this regard, we cannot discard the possible sporadic presence of an additional highly circularly polarized component, possibly disconnected from the main one and driven by coherent mechanisms akin to ECM or plasma emission. If driven by the ECM, the radio emission holds direct information about the magnetic fields at play, with $\nu$ = 2.8 $B$[G] MHz ($\nu$ is the cut-off frequency). This suggests the local magnetic field to be $<6.4$ kG, assuming a cut-off frequency $\sim 18$ GHz (upper limit of Ku band). 

The slightly resolved emission also hints at the possibility that the emission is composed by more than one component: one of them contributing to the thermal emission and the other one responsible for the (partially sporadic) non-thermal component. This can be compatible with both a single or a binary star. The possibility of a radio binary (unresolved so far) would be supported by the fact that the dynamical mass of the source is twice the mass expected from near-IR observations \citep{Zurlo, Alves_2020}. Since the emission patterns from [BHB2007]-1 have been previously seen in other radio binaries \citep{YLW_15_1, YLW_15_2}, this scenario deserves further investigation. A very well known example, with some analogies to our case, is the southern radio source (T Tau Sr) in the T Tauri multiple star system \citep{T_Tauri}. One of this components produces time-variable, circularly polarized radio emission as also observed for two epochs of [BHB2007]-1, while the other component is extended, unpolarized and only moderately variable, with a spectral index typical of optically thin free-free radiation. Past studies have proposed that this variable, polarized emission can have a magnetic origin \citep{T_tauri_2003, T_tauri_Smith}, while the free-free component can be produced by the interaction of a supersonic stellar wind driven by T Tau Sb and dense surrounding material, possibly associated with the circumbinary structure around the pair \citep{T_Tauri}.

Finally, we could not clearly detect the sub-stellar object that is arguably opening up a visible gap in the disk \citep{Alves_2020,Zurlo}, although we see marginal stokes V signals in two of the epochs as explained in Section 3. Apart from that, there also appears a weak Stokes I signal around the proposed position of the sub-stellar object in one of the observations taken on March 30$^{\rm th}$, 2022, but the emission does not coincide exactly with the position of the sub-stellar object. 

In this respect, we suggest that it would be ideal to cover both the low-frequency (GHz and below) and the 50-200 GHz window in the future, to monitor the variability and the overall spectrum of this YSO. In particular, additional data in Stokes V might shed some light on the nature of the polarized component. Finally, it would be very important to search for signatures of the tentative radio companion.

\begin{acknowledgements}
SK, JMG, DV, ASM, OM and FDS's work is partially supported by the program Unidad de Excelencia María de Maeztu, awarded to the Institut de Ciències de l'Espai (CEX2020-001058-M). SK carried out this work within the framework of the doctoral program in Physics of the Universitat Aut\`onoma de Barcelona. SK, DV and OM are supported by the European Research Council (ERC) under the European Union’s Horizon 2020 research and innovation programme (ERC Starting Grant "IMAGINE" No. 948582, PI: DV). JMG acknowleges support by grant PID2020-117710GB-I00(MCI-AEI-FEDER,UE). AZ acknowledges support from ANID -- Millennium Science Initiative Program -- Center Code NCN2021\_080. 
FDS acknowledges support from a Marie Curie Action of the European Union (Grant agreement 101030103). T.B. acknowledges financial support from the FONDECYT postdoctorado project number 3230470. This work has also made use of data from the European Space Agency (ESA) mission {\it Gaia} (\url{https://www.cosmos.esa.int/gaia}), processed by the {\it Gaia}
Data Processing and Analysis Consortium (DPAC, \url{https://www.cosmos.esa.int/web/gaia/dpac/consortium}) 
Funding for the DPAC has been provided by national institutions, in particular the institutions participating in the {\it Gaia} Multilateral Agreement.
\end{acknowledgements}
------------------------------------------------------------
\bibliographystyle{aa}
\bibliography{main}

\onecolumn
\begin{appendix}
\section{Time Series for all the observations}

\begin{figure*}[h!]
\centering

\begin{subfigure}{0.45\columnwidth}
  \centering
  \includegraphics[width=\linewidth, height = 4.75cm]{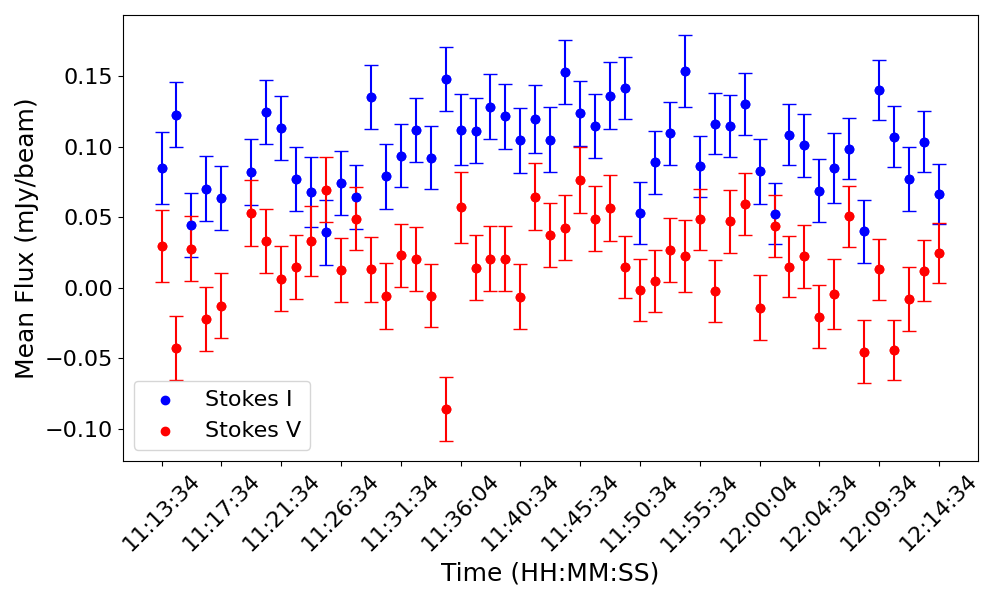}
  \caption{05/03/2022}
\end{subfigure}%
\begin{subfigure}{0.45\columnwidth}
  \centering
  \includegraphics[width=\linewidth, height = 4.75cm]{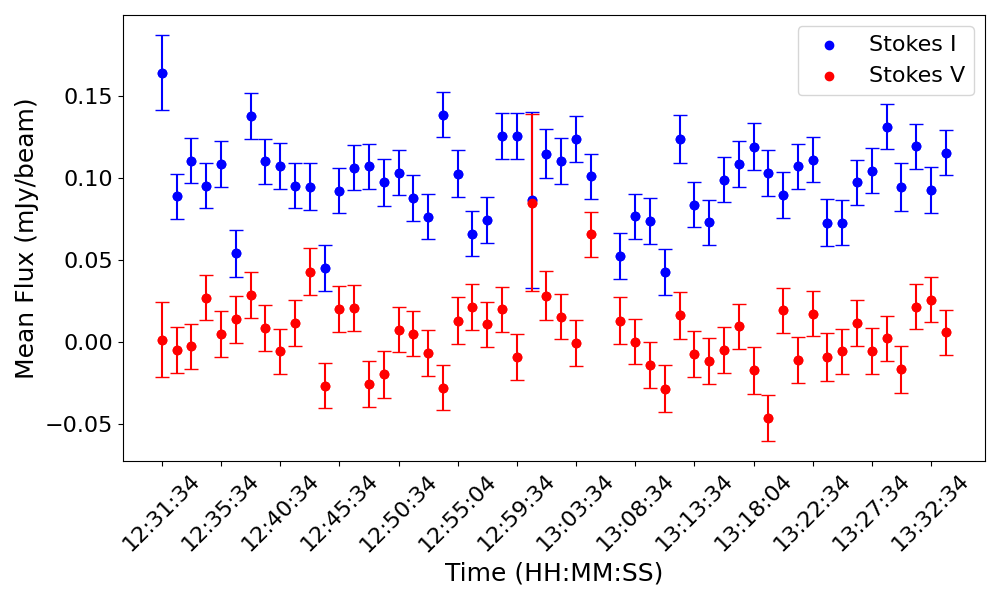}
  \caption{09/03/2022}
\end{subfigure}

\begin{subfigure}{0.45\columnwidth}
  \centering
  \includegraphics[width=\linewidth, height = 4.75cm]{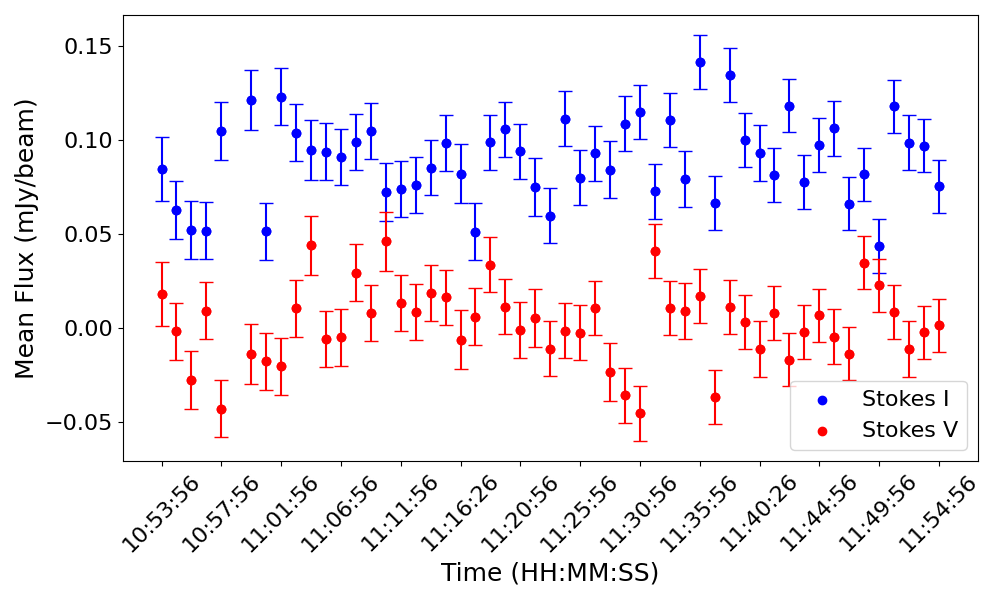}
  \caption{10/03/2022a}
\end{subfigure}%
\begin{subfigure}{0.45\columnwidth}
  \centering
  \includegraphics[width=\linewidth, height = 4.75cm]{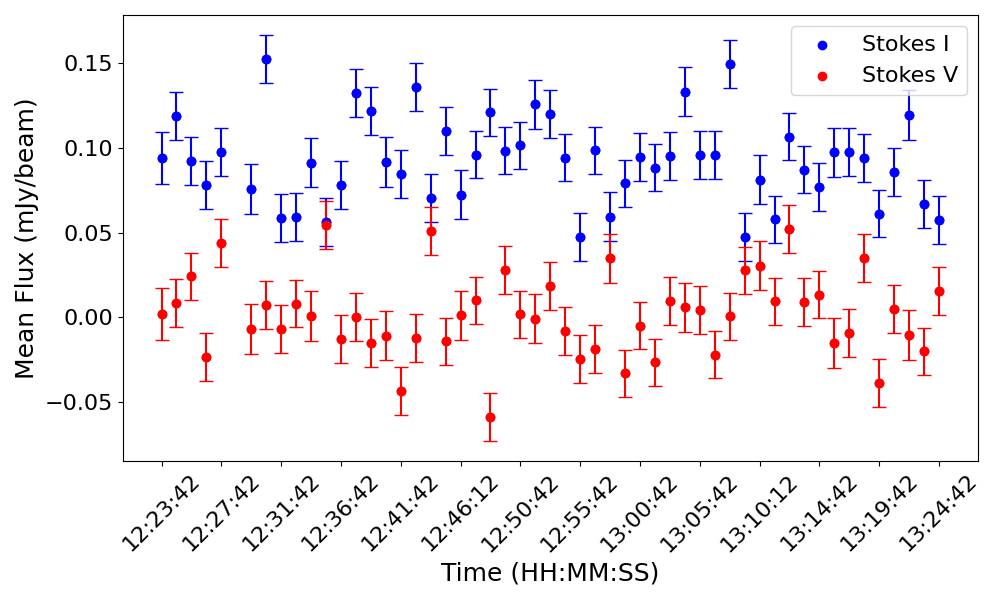}
  \caption{10/03/2022b}
\end{subfigure}

\begin{subfigure}{0.45\columnwidth}
  \centering
  \includegraphics[width=\linewidth, height = 4.75cm]{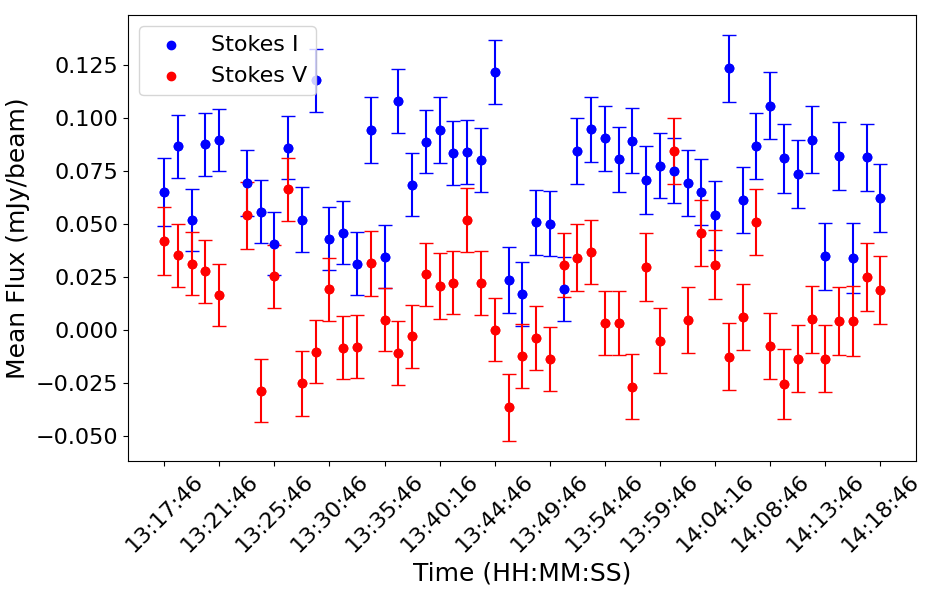}
  \caption{26/03/2022}
\end{subfigure}%
\begin{subfigure}{0.45\columnwidth}
  \centering
  \includegraphics[width=\linewidth, height = 4.75cm]{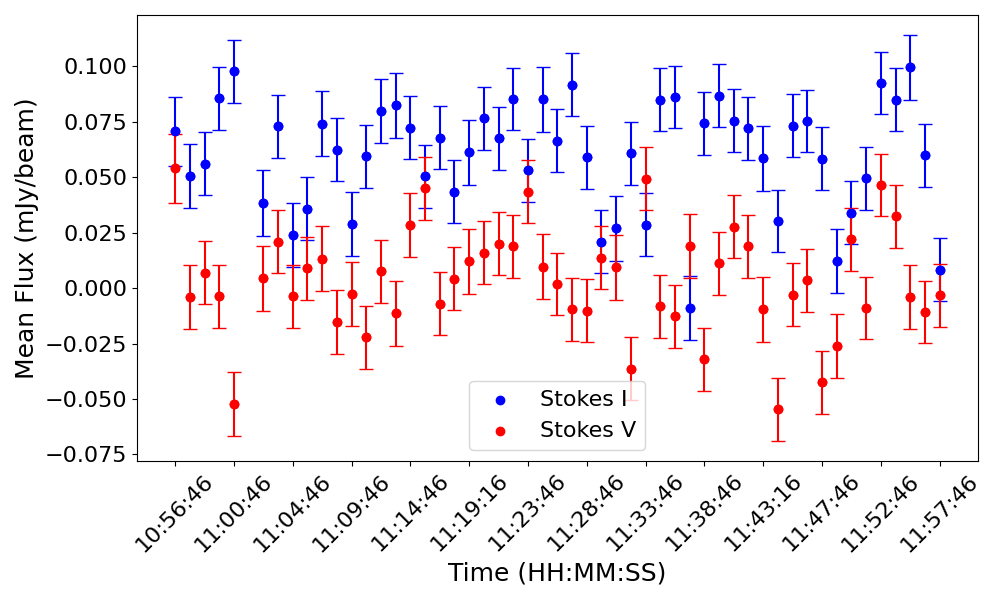}
  \caption{27/03/2022}
\end{subfigure}

\begin{subfigure}{0.45\columnwidth}
  \centering
  \includegraphics[width=\linewidth, height = 4.75cm]{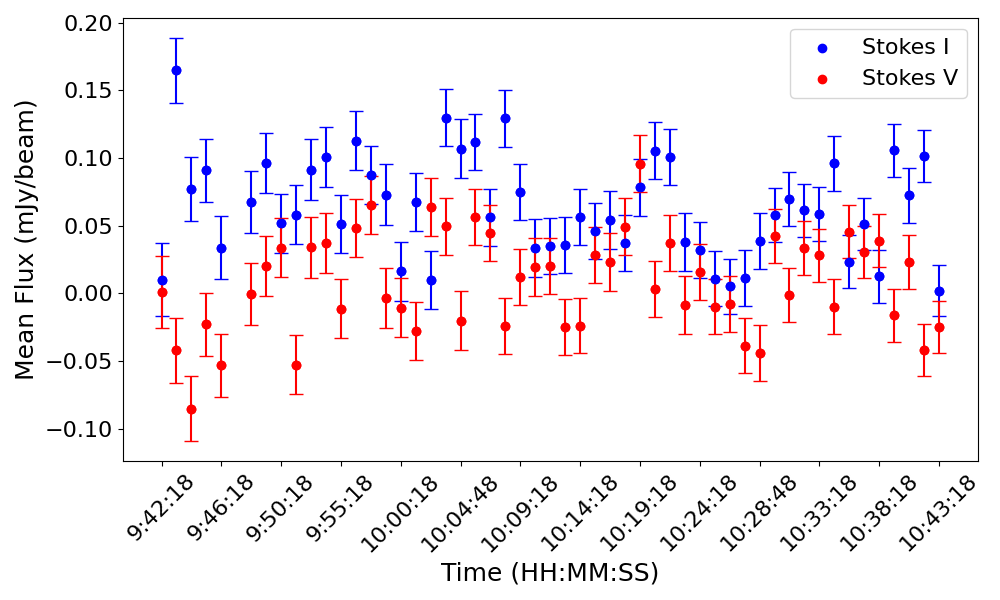}
  \caption{30/03/2022}
\end{subfigure}%
\begin{subfigure}{0.45\columnwidth}
  \centering
  \includegraphics[width=\linewidth, height = 5cm]{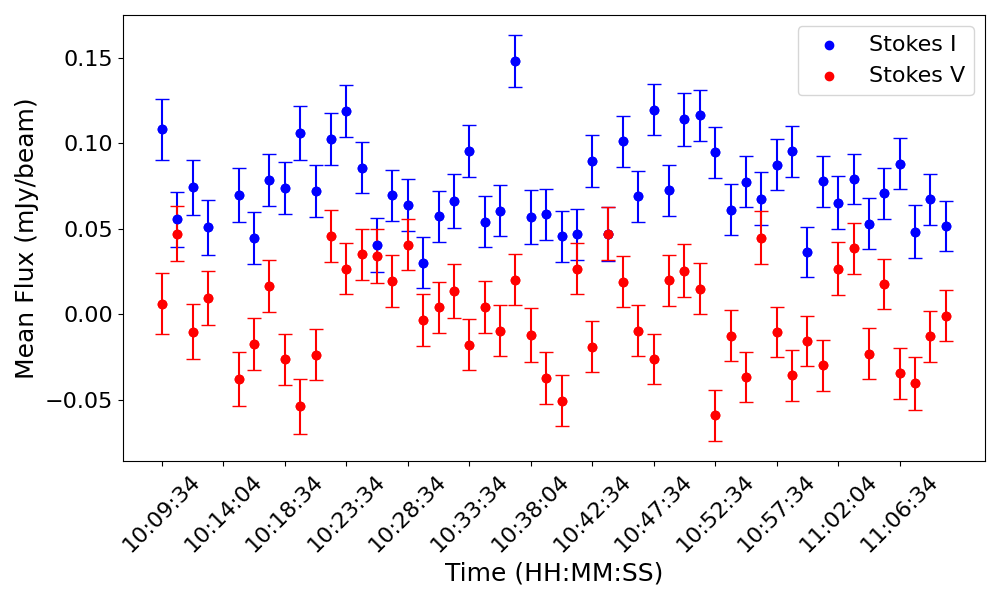}
  \caption{31/03/2022}
\end{subfigure}

\caption{60s time series for all the epochs in Stokes I and Stokes V. These plots have been generated by plotting the real part of amplitude versus time for the target field, using casa task {\tt visstat}. The error bars represent the statistically calculated standard deviation in the data.}
\label{fig:images}
\end{figure*}



\end{appendix}

\end{document}